# Video-rate gigapixel ptychography via space-time neural field representations

Ruihai Wang[1,†], Qianhao Zhao[1,†,*], Zhixuan Hong[1,†], Qiong Ma[1], Tianbo Wang[1], Lingzhi Jiang[1], Liming Yang[1], Shaowei Jiang[1], Feifei Huang[1], Thanh D. Nguyen[1], Leslie Shor[2], Daniel Gage[3], Mary Lipton[4], Christopher Anderton[4], Arunima Bhattacharjee[4], David Brady[5], and Guoan Zheng[1,*]

[1]Department of Biomedical Engineering, University of Connecticut, Storrs, CT 06269, USA
[2]Department of Chemical & Biomolecular Engineering, University of Connecticut, Storrs, CT 06269, USA
[3]Department of Molecular and Cell Biology, University of Connecticut, Storrs, CT, 06269, USA
[4]Pacific Northwest National Laboratory, Department of Energy, Richland, WA, 99354, USA
[5]Wyant College of Optical Sciences, University of Arizona, Tucson, Arizona 85721, USA
[†]These authors contributed equally to this work
[*]Email: qianhao.zhao@uconn.edu; guoan.zheng@uconn.edu

**Abstract:** Achieving gigapixel space-bandwidth products (SBP) at video rates represents a fundamental challenge in imaging science. Here we demonstrate video-rate ptychography that overcomes this barrier by exploiting spatiotemporal correlations through neural field representations. Our approach factorizes the space-time volume into low-rank spatial and temporal features, transforming SBP scaling from sequential measurements to efficient correlation extraction. The architecture employs dual networks for decoding real and imaginary field components, avoiding phase-wrapping discontinuities plagued in amplitude-phase representations. A gradient-domain loss on spatial derivatives ensures robust convergence. We demonstrate video-rate gigapixel imaging with centimeter-scale coverage while resolving 308-nm linewidths. Validations span from monitoring sample dynamics of crystals, bacteria, stem cells, microneedle to characterizing time-varying probes in extreme ultraviolet experiments, demonstrating versatility across wavelengths. By transforming temporal variations from a constraint into exploitable correlations, we establish that gigapixel video is tractable with single-sensor measurements, making ptychography a high-throughput sensing tool for monitoring mesoscale dynamics without lenses.

## Introduction

The space-bandwidth product (SBP) quantifies the total number of effective pixels during acquisition, defining the intrinsic information limit of imaging systems[1,2]. Achieving gigapixel-scale SBP at video rates faces two fundamental barriers: optical systems must collect and relay gigapixel-scale information, while sensors must capture and readout this information at camera framerates. Conventional lenses suffer from aberrations that grow polynomially with both aperture size and field angle. This scaling law forces a trade-off: high numerical aperture (NA) objectives minimize aberrations but cover only a small field of view, while wide-field lenses attempting large field coverage accumulate aberrations that degrade resolution. It is challenging to achieve both wide field coverage and high resolution for delivering gigapixel SBP. Similarly, image sensors face physical constraints on pixel count, size, and readout speed that prevent gigapixel acquisition at video rates. Together, these optical and sensor SBP limitations create a seemingly insurmountable barrier to video-rate gigapixel imaging.

Current computational imaging methods overcome the static SBP limit through sequential acquisition strategies. Whole slide scanners mechanically tile high-resolution images across real space[3,4,5,6], while Fourier ptychography synthesizes apertures in reciprocal space through angle-varied illumination[7,8]. Both achieve gigapixel SBP but require minutes of dataset acquisition. These approaches fundamentally treat gigapixel SBP as a multiplexing problem, where acquisition time scales linearly with real- or reciprocal-space coverage. Coded aperture temporal imaging offers an alternative temporal multiplexing strategy by mechanically translating coded masks during exposure to compress video sequences into snapshot measurements[9]. However, this approach addresses temporal compression without achieving the resolution improvement necessary for gigapixel SBP, thereby exchanging one bottleneck for another. In contrast to sequential multiplexing, multiscale optical designs employ a large objective lens that collects light from a wide angle, then relay this to an array of smaller lenses with detectors that each image a portion of the field[10]. This architecture demonstrates that lens speed and field of view can be scale-independent, enabling snapshot gigapixel photography[11,12]. Likewise, array microscopy pursues the same parallel acquisition goal through distributed imaging systems and requires proportionally complex multi-camera hardware[6, 13, 14, 15]. Whether through time-multiplexed sequential acquisition or hardware-intensive parallelization, the fundamental constraint remains: achieving gigapixel SBP demands either sacrificing temporal resolution or dramatically increasing system complexity, making single-sensor gigapixel video acquisition seemingly impossible.

Here we demonstrate video-rate gigapixel ptychography using space-time neural fields to address the traditional SBP scaling challenges. Neural field representations parameterize continuous space-time volumes through coordinate-based networks[16, 17, 18, 19, 20, 21, 22, 23, 24, 25, 26, 27, 28, 29, 30], efficiently encoding high-dimensional data while naturally regularizing inverse problems. By representing dynamic scenes as continuous neural fields, we can exploit the rich spatiotemporal correlations inherent in videos. For addressing the optical SBP challenge, we employ a coded surface that maintains gigapixel-scale information capacity without aberration issues of conventional lenses. For addressing the sensor SBP challenge, our neural fields distribute the bandwidth burden across time, where each measurement simultaneously enhances spatial resolution toward gigapixel scales while maintaining video-rate temporal sampling. This fundamentally changes how SBP scales -- SBP grows with correlation extraction rather than increasing system complexity or time.

Our system resolves 308-nm linewidths across centimeter-scale fields at 30 frames per second (fps), establishing that single-sensor gigapixel video is achievable through properly co-designed optics, sensing scheme, and computation. Importantly, the approach generalizes across wavelengths, successfully recovering time-varying probe beams in extreme ultraviolet (EUV) experiments where wavefront instabilities challenge conventional methods[31]. It also naturally extends to electron[32, 33], X-ray ptychography[34], where radiation-induced sample evolution presents analogous spatiotemporal challenges requiring robust dynamic phase retrieval[35].

## Results

**Video-rate ptychography with physics-informed neural fields**

Figure 1 shows the video-rate ptychography framework and its experimental validation. The architecture (Fig. 1a, top panel) factorizes the space-time volume into hash-encoded spatial features[19] and compact temporal features[20, 22, 26], compressing gigabytes of data into megabyte representations without involving motion models. To efficiently represent spatial coordinates, we employ multi-resolution hash encoding with $L$ levels and $F$ features per level (Methods). At any spatiotemporal point, these hash-encoded spatial features ($w \times h \times n$ channels) combine with temporal features through Hadamard product before feeding dual multi-layer perceptrons (MLPs) that decode real and imaginary field components separately. We adopt this real-imaginary parameterization rather than conventional amplitude-phase representations to maintain continuous, smooth functions throughout the reconstruction volume, avoiding the non-differentiable $2\pi$ discontinuities at phase-wrapping boundaries that trap optimization in local minima. This continuity ensures stable gradient flow during optimization, enabling successful convergence even for samples with large phase variations where amplitude-phase methods produce artifacts and convergence failures (Supplementary Fig. S1a-d vs. e-h).

The optimization pipeline (Fig. 1a, bottom panel) employs a gradient-domain loss[23, 36, 37, 38, 39] operating on spatial derivatives of amplitude differences in real space: $\left\| \nabla \left( \sqrt{J_t(x,y)} - \nabla \sqrt{I_t(x,y)} \right) \right\|_2$, where $J_t(x,y)$ denotes the measured intensity at time $t$, $I_t(x,y)$ denotes the estimated intensity from the forward imaging model of ptychography, $\nabla$ denotes the gradient operator, and $\|\cdot\|_2$ denotes the $L_2$ norm. This gradient-domain formulation departs from conventional absolute amplitude losses, providing enhanced robustness for dynamic phase reconstruction. The loss back-propagates through the dual MLPs to update both network weights and the space-time feature tensors. The physics-informed approach ensures that learned features respect optical constraints while requiring no data pre-training. The synergistic combination of low-rank feature decomposition, gradient-domain loss, and the real-imaginary formulation enables robust convergence for challenging dynamic samples with rapid phase variations exceeding multiple wavelengths -- conditions where conventional iterative phase retrieval methods fail.

Our experimental configuration (Fig. 1b) employs a distinct measurement strategy compared to conventional or Fourier ptychography where confined measurements are acquired in either real or reciprocal space. Dynamic specimens are illuminated by a laser diode without additional optics, and we perform image acquisition by translating a coded sensor beneath the dynamic object using voice coil actuators. The coded surface on the image sensor, featuring randomized intensity and phase scatters, serves as an effective probe function as in conventional ptychography[6, 35]. The system calibration and positional tracking process are detailed in Supplementary Note 1. At each coded sensor position, the system generates a unique diffraction pattern through modulation of the object's exit wave field. Each of these measurements encodes full-field spatial information across all frequencies, contrasting with sequential confined sampling as in conventional or Fourier ptychography. The imaging model and conventional phase retrieval strategies are detailed in Supplementary Note 2.

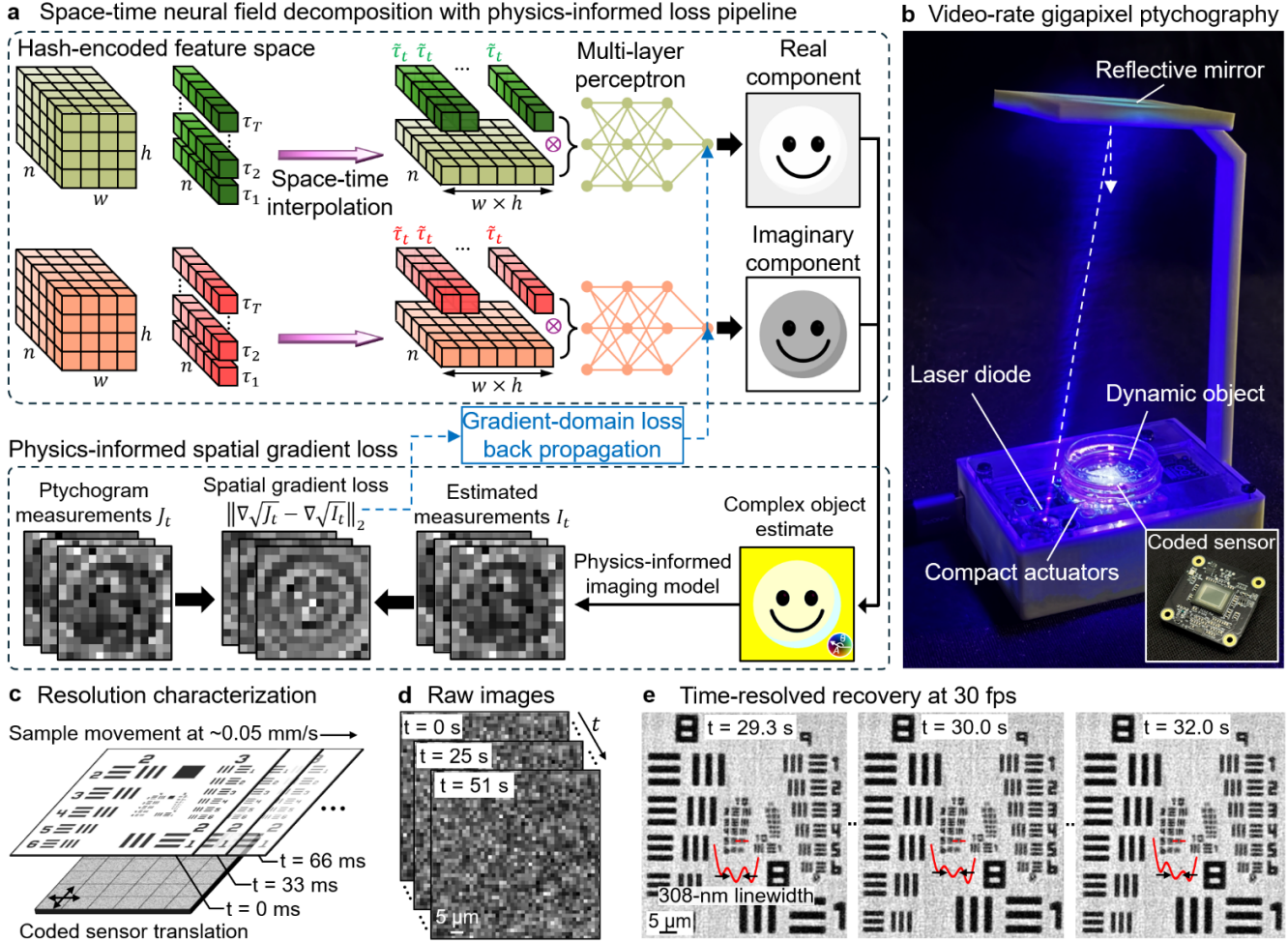

**Fig. 1| Space-time neural fields for video-rate gigapixel ptychography. a**, Top panel: Neural field architecture showing low-rank decomposition of space-time volume into spatial and temporal feature tensors. The temporal basis features $\tau_i$ ($i = 1 \cdots T$) define a compact representation of temporal variation, from which continuous temporal features $\tilde{\tau}_t$ are interpolated for arbitrary time indices. The spatial feature tensor is multi-resolution hash-encoded from spatial coordinates into feature vectors and reshaped into a two-dimensional array. Each spatial feature vector then performs a Hadamard product with the same temporal feature $\tilde{\tau}_t$. Each combined feature vector is fed into dual MLPs that output the real and imaginary components of the complex field at that spatial coordinate, and all outputs together form a reconstructed frame at time $t$. Bottom panel: Physics-informed optimization pipeline adopts the gradient-domain loss to update the MLPs' weights and feature tensors. **b**, Experimental setup where a blue laser diode illuminates the dynamic object. A coded sensor is translated by compact actuators for diffraction data acquisition. Inset shows the coded sensor chip with the coded surface serving a spatially extended ptychographic probe beam. **c**, Resolution characterization of a USAF target moving at ~0.05 mm/s. **d**, Raw diffraction image stack captured during target translation. **e**, Time-resolved reconstruction resolving the 308-nm linewidth of the group 10 element 5.

In Fig. 1c, we test the resolution performance by translating a USAF target at ~0.05 mm/s. We acquire the raw images for a duration of 50-second, at 30 frames per second (Fig. 1d). Critically, the neural field framework leverages the entire 50-second dataset to jointly recover both spatial and temporal features. As shown in Fig. 1e, we resolved the 308-nm linewidth at different time points (detailed in Supplementary Note 3). This collective optimization strategy is essential: attempting reconstruction from a single (or a few) diffraction pattern yields an underdetermined system that cannot converge. By simultaneously processing all temporal data through the space-time factorization, our framework exploits redundancy across time points to achieve robust convergence for dynamic samples.

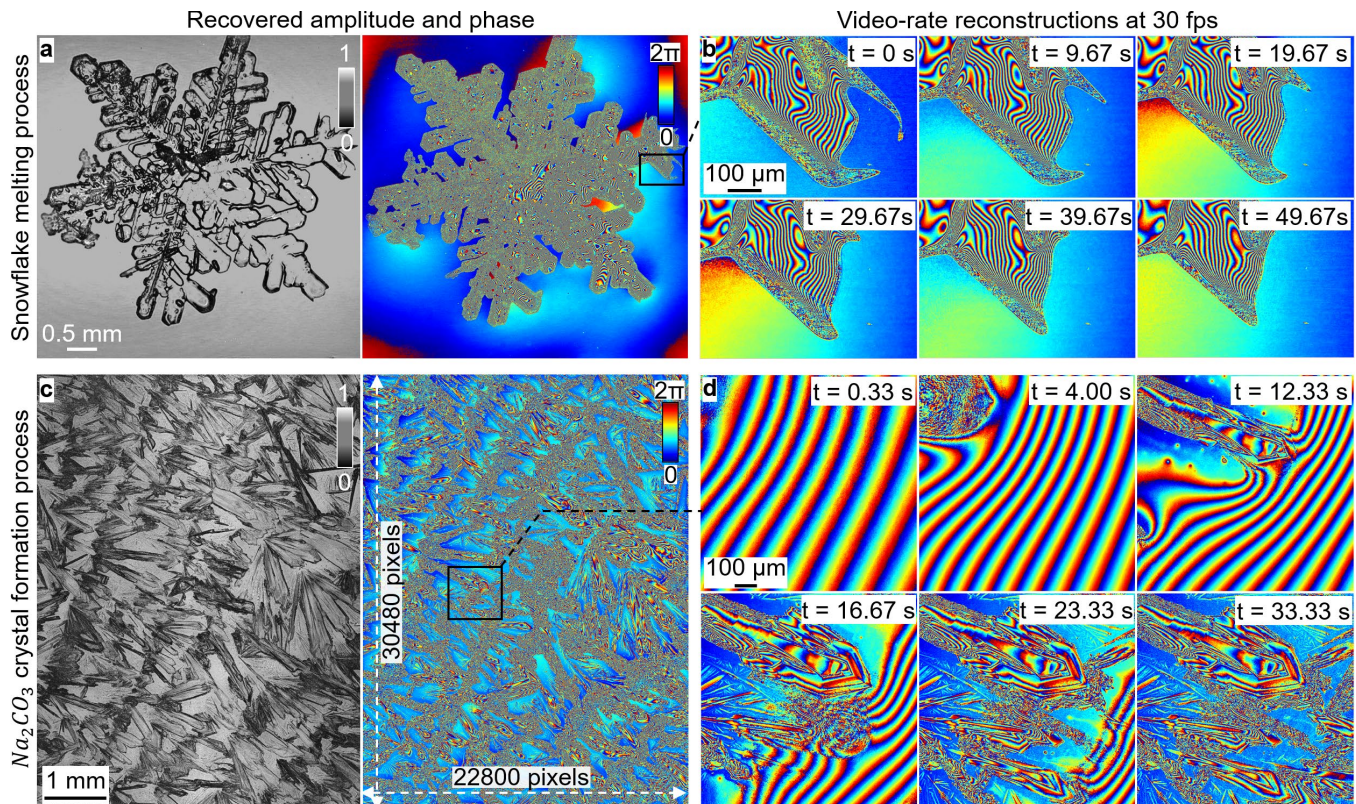

**Fig. 2| Video-rate, gigapixel imaging of dynamic crystallization processes enabled by joint spatiotemporal optimization of neural fields. a**, Recovered amplitude (left) and phase (right) of the snowflake melting process, showing preserved dendritic structure across a large field of view. **b**, Time-lapse phase imaging over the 50-second duration, revealing melting front propagation at selected time points. **c**, Gigapixel-scale crystal formation process over the entire ~40 mm$^2$ sensor field of view. **d**, Temporal sequence showing crystal formation process at selected time points. The reconstruction leverages all temporal data through space-time neural fields, enabling artifact-free high-resolution imaging impossible with conventional frame-by-frame approaches. Supplementary Fig. S2 and Movies 1-2 show the melting processes of large and small snowflakes. Supplementary Figs. S3-S4 and Movie 3 show the recovered gigapixel-scale video of the crystal formation process.

**Video-rate gigapixel imaging at the mesoscale**

Snowflake imaging has fascinated microscopists since Wilson Bentley captured the first photograph in 1885, establishing that no two snowflakes are identical. Despite over a century of technical advancement of microscopy, no existing approach simultaneously achieves centimeter-scale coverage, sub-micrometer resolution, and video-rate temporal sampling necessary to capture melting dynamics of fragile ice crystals. The reported framework overcomes these fundamental limitations through factorizing the entire measurement sequence into shared spatial and temporal features, exploiting temporal redundancy to regularize the ill-posed dynamic phase retrieval problem. Figure 2a-b, Supplementary Fig. S2, and Supplementary Movies 1-2 demonstrate this capability across snowflakes of varying sizes and morphologies. The recovered phase maps quantitatively track melting front propagation and reveal evolution of water film thickness. Fine dendritic structures remain clearly resolved throughout the thermal transition, demonstrating the framework's ability to maintain spatial fidelity during morphological changes. Complementary to melting dynamics, crystallization from solution presents phase accumulation challenges as material thickness rapidly increases. We demonstrate video-rate gigapixel monitoring of crystal formation across the entire sensor field of view, capturing the complete evolution from dendritic expansion to polycrystalline film formation at unprecedented scale (Fig. 2c-d). This gigapixel-level reconstruction simultaneously resolves sub-micron crystalline features while maintaining centimeter-scale coverage (Supplementary Figs. S3-S4).

Critical to imaging performance of video-rate ptychography is the synergistic choice of loss function and optical field representation. We demonstrate this using the crystallizing sample where rapid phase accumulation creates optical path differences exceeding many 2π wraps. In Supplementary Fig. S5a-d, gradient-domain loss with real-imaginary representation achieves successful convergence with artifact-free reconstruction. Other choices fail to converge or exhibit significant artifacts. We also compare the reported neural field framework against common iterative phase retrieval algorithms including ePIE[40], momentum-PIE (mPIE)[41], Quasi-Newton[42], and least-squares[43] methods (Supplementary Fig. S5e-h). All conventional methods exhibit artifacts and convergence failures. Supplementary Movie 4 further compares the reconstructed videos between the reported framework and other common iterative phase retrieval methods. This comparison reveals that frame-by-frame reconstruction is fundamentally underdetermined for dynamic imaging. In contrast, our space-time neural field jointly optimizes across all temporal measurements, exploiting spatiotemporal correlations through shared low-rank features to transform the severely underdetermined problem into a tractable one. Supplementary Fig. S6 further confirms these advantages with a microneedle patch, where the reported framework demonstrates superior reconstruction quality.

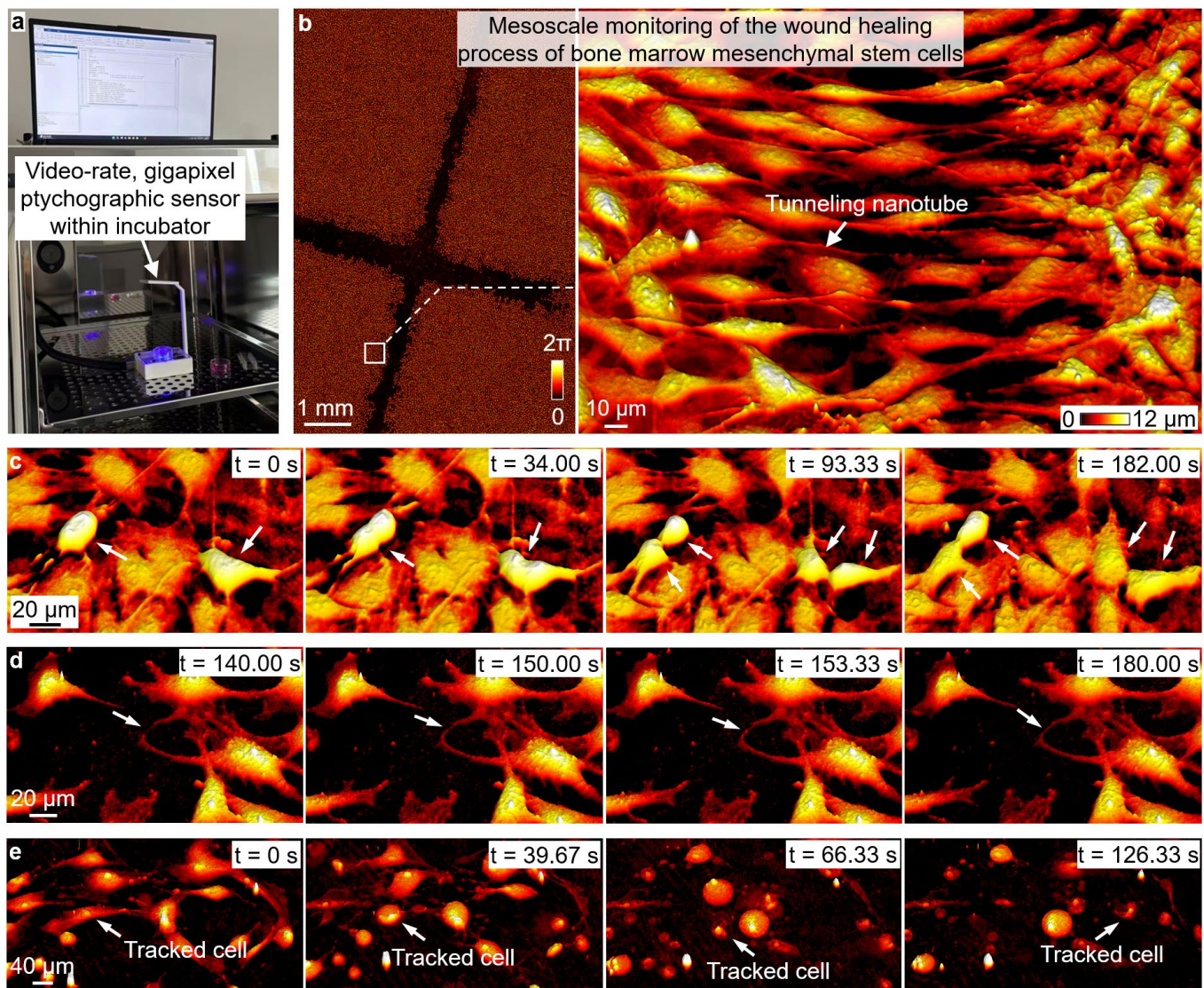

**Fig 3| Live stem cell monitoring within an incubator. a**, Experimental setup showing the potable video-rate ptychographic sensor in an incubator. **b**, Reconstructed mesoscale height map showing wound healing process of the bone marrow mesenchymal stem cells. The white arrow highlights the recovered tunneling nanotube with high spatial resolution. **c**, Cell division sequence showing mitotic rounding, cytokinesis, and daughter cell separation. **d**, Intercellular transport through tunneling nanotubes. **e**, Cell necrosis showing rapid swelling followed by membrane rupture. Supplementary Movies 5-7 show the videos of cell division, intercellular transport, and necrosis. Supplementary Figs. S7-S8 and Movie 8 show the post-measurement refocusing capability.

**Live-cell dynamics monitoring with post-measurement digital refocusing**

Video-rate ptychography enables label-free monitoring of cellular processes by detecting quantitative phase variations without the phototoxicity associated with fluorescent labeling. To demonstrate this capability, we monitored bone marrow mesenchymal stem cells during wound healing assays within an incubator environment (Fig. 3a). Our mesoscale gigapixel phase reconstructions reveal detailed cellular morphology and intercellular structures across the entire sensor field of view, with the recovered height maps resolving tunnelling nanotubes that form critical communication pathways between cells (Fig. 3b). The high temporal resolution captures complete cell division dynamics, revealing characteristic mitotic rounding accompanied by progressive phase shifts that precede cytokinesis (Fig. 3c). In Fig. 3d, our system can also visualize intercellular transport along tunnelling nanotubes -- events typically challenging to observe without fluorescent markers. In Fig. 3e, we induced necrosis through hypotonic shock and the resulting time-lapse sequences capture rapid cellular swelling followed by catastrophic membrane rupture, enabling quantitative monitoring of cell death kinetics.

A distinctive advantage of video-rate ptychography is its post-acquisition digital refocusing capability -- essential for dynamic samples where focal planes vary during acquisition. Unlike conventional and Fourier ptychography, which require explicit modeling of object-probe interactions, our system reconstructs the optical field exiting the sample. This fundamental difference enables digital propagation of the recovered complex field to arbitrary axial positions post-measurement, without the thin-sample approximations that constrain traditional ptychographic approaches[44, 45]. Supplementary Fig. S7 validates this capability by computationally refocusing microneedles and live stem cell cultures, revealing how different structural features achieve optimal focus at distinct depths. Supplementary Fig. S8 compares fixed-plane reconstructions with dynamically refocused sequences during necrosis, demonstrating superior image quality through adaptive focusing. Supplementary Movie 8 shows the post-measurement refocusing process of a microneedle patch. By eliminating real-time autofocus requirements, the post-acquisition flexibility simplifies experimental protocols while enabling retrospective 3D analysis of archived datasets.

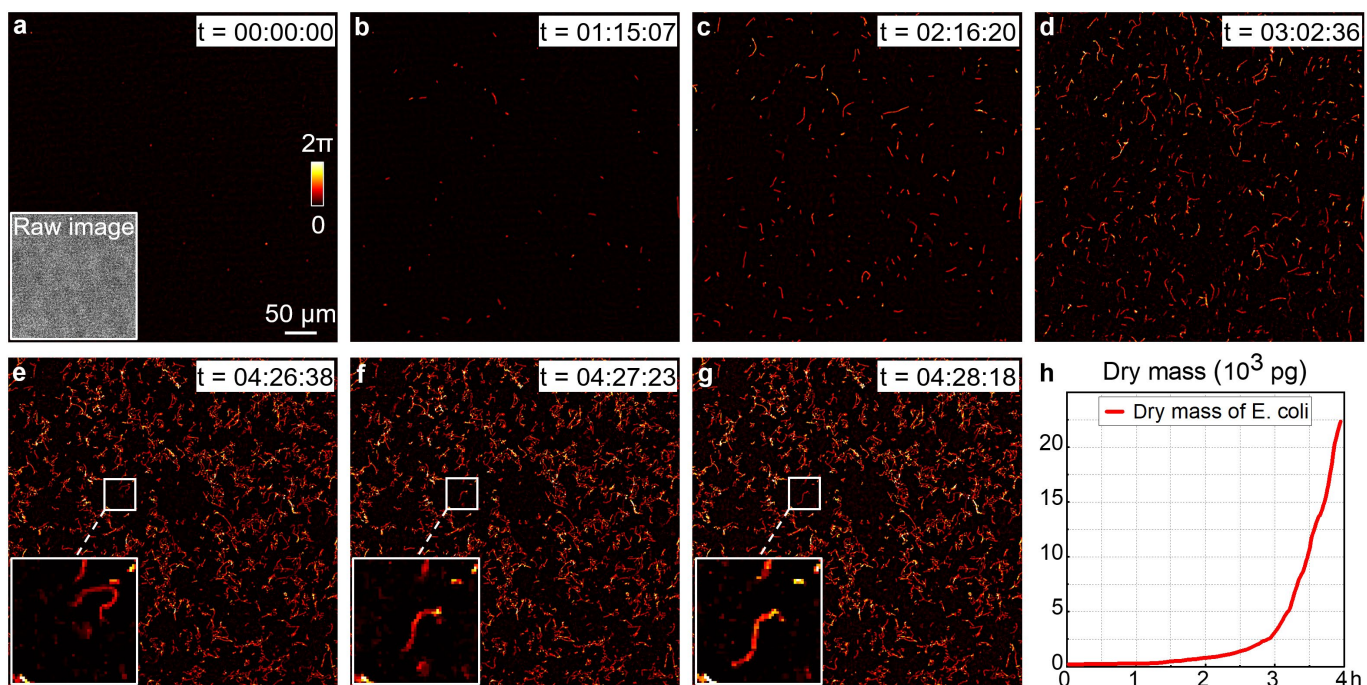

**Fig. 4| Quantitative monitoring of bacterial growth dynamics in liquid culture. a-g**, Time-lapse phase imaging of *E. coli* proliferation within a free-flow microfluidic channel. Gentamicin added in the culture medium induces filamentation, causing some cells to elongate without dividing -- a behavior associated with antibiotic resistance and biofilm formation. Magnified regions in (e-g) highlight individual cells. **h**, Quantitative dry mass accumulation shows exponential growth. The measurement enables label-free growth kinetics quantification in flowing liquid culture with picogram sensitivity. Supplementary Movie 9 shows the real-time bacterial growth process in liquid culture. Supplementary Movie 10 shows the real-time bacterial growth on solid agar plate.

In Fig. 4, we also demonstrate the platform's versatility through longitudinal tracking of bacterial growth with implications for antibiotic susceptibility testing (AST). Time-lapse phase imaging captures *E. coli* proliferation within a free-flow microfluidic channel, where antibiotic gentamicin treatment induces filamentation – some cells elongate while continuing to grow. This morphological response, visible as extended thread-like cells in Fig. 4c-g, represents a survival mechanism that protects bacteria from antibiotics and promotes virulence factors including biofilm formation. Our platform tracks both this elongation phenotype and biomass accumulation from sparse inoculation to dense coverage (Fig. 4a-g and Supplementary Movie 9). The quantitative phase signal provides direct biomass measurements, with phase shifts linearly correlating to cellular dry mass density (Methods). Despite cells elongating into filaments rather than dividing normally, we observe exponential mass accumulation in Fig. 4h, revealing how bacteria maintain growth even under antibiotic stress. This label-free quantification could accelerate clinical decision-making by detecting resistance phenotypes hours before conventional culture methods.

Complementing liquid culture monitoring, we extended our approach to solid-phase bacterial growth on agar plates, achieving centimeter-scale field-of-view while maintaining high temporal resolution (Supplementary Fig. S9 and Movie 10). This configuration monitors colony formation and expansion on nutrient agar surfaces with high phase sensitivity. The reported platform thus bridges multiple culture formats, from microfluidic liquid cultures to traditional agar plates, providing continuous, quantitative monitoring across different growth conditions without phototoxicity constraints.

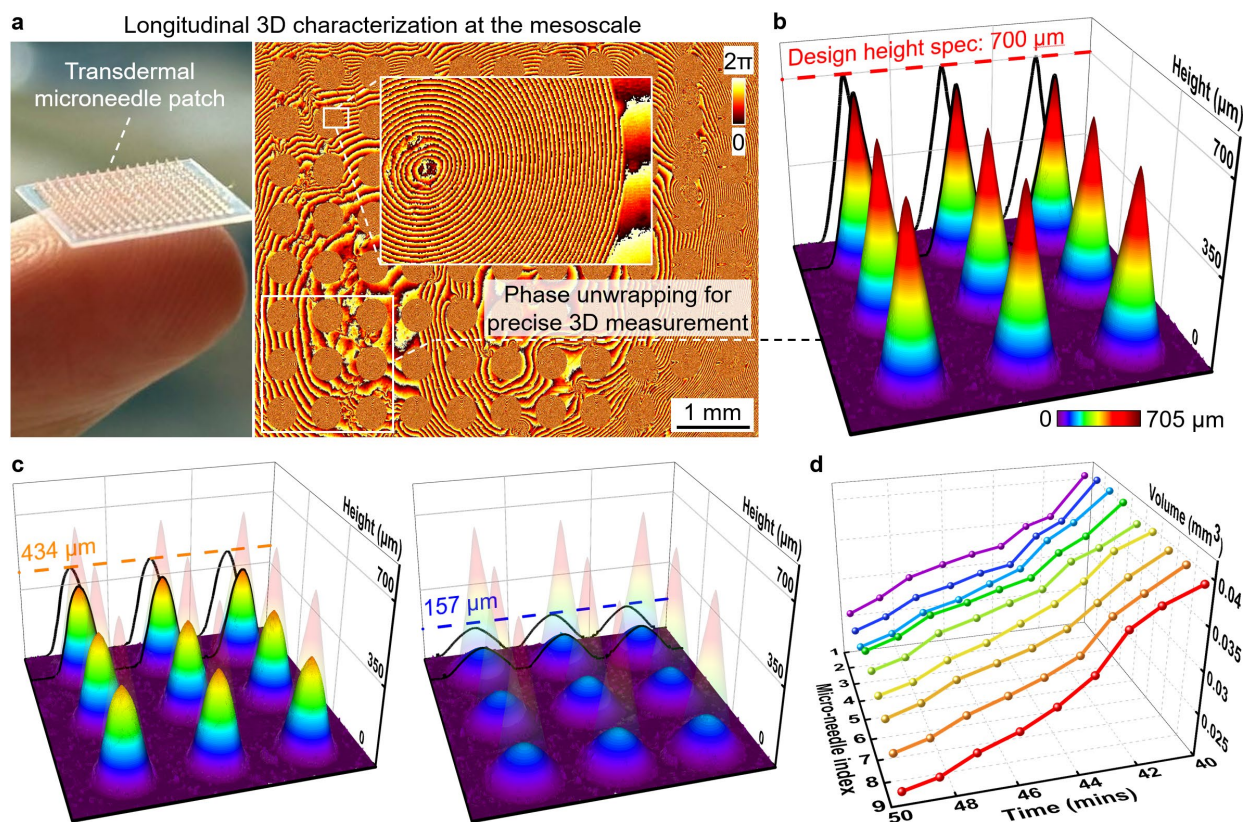

**Fig. 5| Longitudinal 3D characterization of microneedle patch. a**, Transdermal microneedle patch and its corresponding phase map covering a centimeter-scale field of view. The zoomed-in view shows the wrapped phase for height extraction. **b**, 3D reconstruction validating the 700-μm design height with sharp tip geometry preserved across hundreds of needles. **c**, Time-resolved dissolution monitoring showing progressive height reduction. **d**, Graph quantifies dissolution kinetics through volumetric measurements, enabling pharmacokinetic modeling of drug release profiles. Supplementary Fig. S10 shows mesoscale characterization of the microneedle patch. Supplementary Movie 11 shows the dissolution dynamics.

**Longitudinal 3D characterization of biomedical devices**
Transdermal microneedles enable drug delivery through painless self-administration, yet their clinical translation requires standardized quality control methods currently absent from pharmaceutical practice. We demonstrate how video-rate ptychography addresses these unmet needs through non-destructive 3D characterization of microneedle arrays in Fig. 5. The recovered phase maps in Fig. 5a cover the entire centimeter-scale microneedle patch containing hundreds of individual needles. Phase unwrapping converts the wrapped phase to absolute height measurements, confirming 700-μm needle heights matching design specifications in Fig. 5b. The 3D surface reconstruction also shows uniform needle geometry with sharp tips essential for skin penetration, validating manufacturing consistency across the array. Supplementary Fig. S10 shows the mesoscale 3D characterization process. In Fig. 5c and Supplementary Movie 11, we perform time-resolved imaging to capture dissolution dynamics as microneedles interact with surrogate dissolution medium. Sequential 3D reconstructions track volumetric reduction, where individual needles show slightly varying dissolution rates potentially reflecting local flow conditions or manufacturing variations. The volumetric data enables precise pharmacokinetic modeling, with dissolution curves extracted for each needle providing statistical distributions of release rates -- essential data for regulatory approval. While conventional quality control requires destructive sampling at multiple timepoints, our approach provides continuous, non-invasive monitoring at video rates, transforming pharmaceutical characterization from endpoint testing to real-time validation.

**Characterizing time-varying probes in extreme ultraviolet ptychography**
To validate our framework beyond visible light, we tested it on EUV ptychography where time-varying illumination presents severe reconstruction challenges[31]. In Fig. 6, we demonstrate that the reported space-time neural fields achieve robust reconstruction even with reduced measurements and minimal iterations -- capabilities essential for EVU, X-ray and electron imaging where source stability and radiation damage are limiting factors.

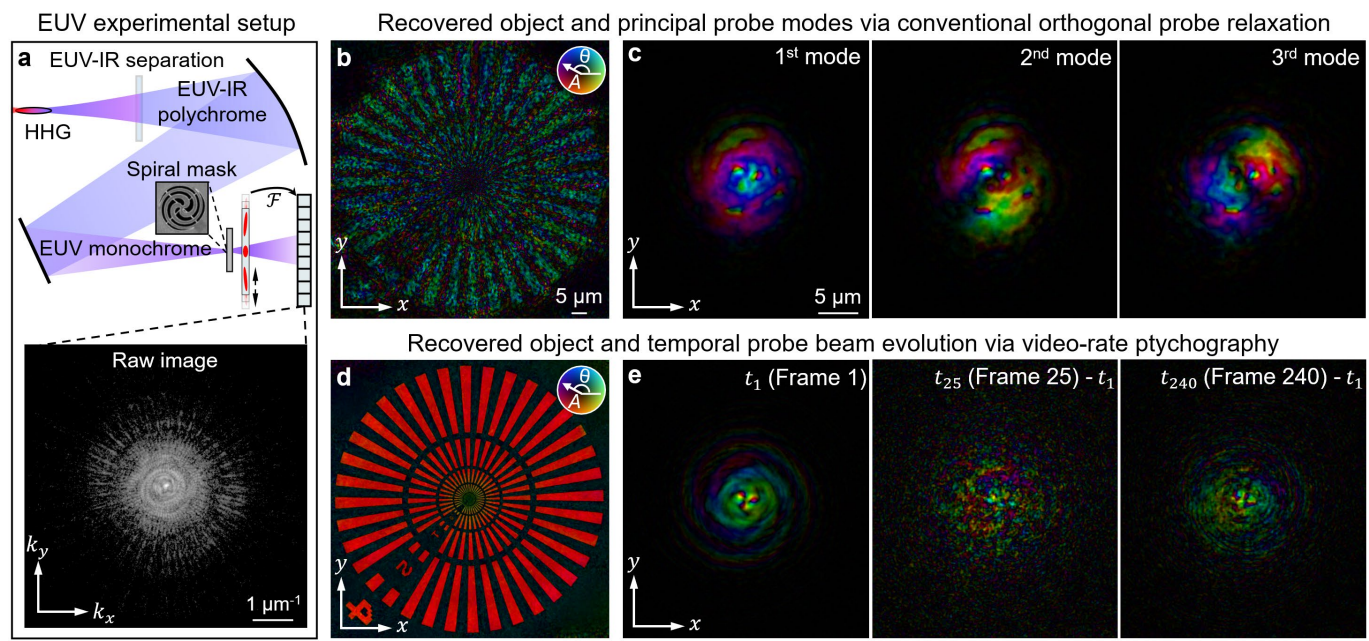

**Fig. 6| Time-varying probe characterization in extreme ultraviolet ptychography. a**, Experimental setup showing a spiral phase mask for generating the probe beam on the resolution target, with the raw image captured in reciprocal space (250 frames in total, with 24% overlap in-between adjacent measurements). **b**, Sample reconstruction via conventional orthogonal probe relaxation using 1500 iterations. **c**, The recovered first three principal modes of the varying probe beam. **d**, Sample reconstruction via the reported video-rate ptychography framework using 40 iterations. **e**, The recovered temporal evolution of probe beam across all frames. Supplementary Note 4 show detailed comparisons with conventional methods at 1.2%, 10.3% and 24.4% overlap ratios. Supplementary Movie 12 shows complete time evolution of the recovered probe beam.

Figure 6a shows the EUV experimental setup using a spiral mask for probe generation and a resolution target as the sample[31]. To test robustness under challenging conditions, we used a set of 250 raw diffraction patterns with only ~24% overlap ratio in Fig. 6 -- significantly below the typical 50% overlap in conventional reconstruction. This reduced overlap ratio combined with probe variations during acquisition creates a demanding test case for ptychographic algorithms. As shown in Fig. 6b-c, conventional orthogonal probe relaxation (OPR) methods[46] struggle with these conditions. Figure 6b shows the OPR-reconstructed object with 1500 iterations, exhibiting artifacts and incomplete convergence. The recovered orthogonal modes in Fig. 6c attempt to capture probe variations but produce ambiguous decompositions. In contrast, our video-rate ptychography framework achieves successful reconstruction in 40 iterations (Fig. 6d). Rather than decomposing into orthogonal modes, our framework treats the probe beam as a temporal function across all 250 frames. Figure 6e shows the temporal evolution of the probe beam across different frames. This natural parameterization through temporal features provides stronger constraints than orthogonal decomposition, enabling robust reconstruction from sparse, unstable measurements. We further tested reconstruction limits by reducing overlap ratios to extreme conditions (Supplementary Note 4). At 10.3% overlap, the framework maintains successful object recovery with probe variations. Even at 1.2% overlap, the recovered object remains recognizable despite increased artifacts, demonstrating robustness against undersampling conditions. Supplementary Note 4 provides detailed comparisons across multiple algorithms and conditions. The success with challenging EUV data demonstrates our framework's potential for electron, X-ray, and EUV ptychography, where minimizing exposure while handling beam variations remains critical[47].

## Discussion

Ptychography reconstructs complex optical fields from overlapping diffraction measurements[35, 48, 49], achieving unmatched sensing capabilities across scales -- from atomic-resolution electron microscopy[33], nanoscale X-ray ptychography[34], to meter-scale macroscopic Fourier ptychography[50]. Whether the confined measurement constraint manifests in real space[51] or reciprocal space[7, 8], ptychography faces an irreconcilable conflict: the data redundancy essential for phase retrieval necessitates sequential measurements with substantial overlaps that cannot capture instantaneous dynamic states. This fundamental bottleneck has limited ptychography to quasi-static samples despite its unique ability to surpass physical limitations of lens-based optics. In electron and X-ray ptychography, this limitation becomes critical as radiation damage leads to structural changes during data acquisition, creating an inherent race against sample degradation[35, 47, 48].

The reported sensing scheme addresses these limitations by exploiting spatiotemporal correlations inherent in dynamic scenes. Rather than multiplexing in space or frequency alone, we distribute the bandwidth burden across time through neural field factorization. The key insight is that natural dynamic objects exhibit strong correlations -- spatial structures persist across frames while temporal dynamics follow predictable patterns. By encoding these correlations through low-rank feature decomposition, each measurement simultaneously contributes to spatial resolution enhancement and temporal evolution tracking. This fundamentally changes the scaling relationship: SBP grows with the efficiency of correlation extraction rather than the number of independent measurements.

The physics-informed architecture amplifies this efficiency through synergistic design choices. Gradient-domain optimization exploits spatial-gradient sparsity, enable robust phase retrieval compared to regular amplitude-based optimization. Real-imaginary parameterization maintains smooth optimization landscapes, avoiding phase-wrapping discontinuities. Hash-encoded spatial features provide memory-efficient multi-resolution representation, while temporal factorization captures dynamics without restrictive motion models. These mechanisms enable reconstruction of gigapixel video streams at camera framerates.

Our diverse demonstrations validate capabilities extending from materials science to biomedicine. Time-resolved imaging of snowflake crystallization exemplifies the unique convergence of gigapixel field coverage and video-rate streaming -- a combination unattainable with existing techniques. We simultaneously track dendritic growth patterns across centimeter-scale fields at sub-micron resolution and 30 fps. This enables observation of both microscopic events and macroscopic pattern formation within the same dataset. In biomedical applications, label-free monitoring of stem cell dynamics revealed subcellular structures like tunnelling nanotubes typically requiring fluorescent labeling, while pharmaceutical device characterization recovered 3D shapes that are challenging to get with conventional approaches. The post-acquisition refocusing capability proves particularly valuable for dynamic samples where focal planes shift unpredictably, as observed during cell necrosis. Unlike conventional and Fourier ptychography that model object-probe interactions, our direct field reconstruction enables digital propagation without thin-sample approximations.

The reported sensing scheme also addresses fundamental challenges across ptychographic modalities. In EUV ptychography, source instabilities cause probe variations between acquisitions[31]. We demonstrated successful recovery of both sample structure and time-varying probe evolution, treating illumination fluctuations as continuous temporal functions rather than discrete orthogonal modes. The framework's robustness proves particularly valuable for short-wavelength regimes where minimizing exposure is critical. For X-ray and electron ptychography, radiation damage progressively alters samples during measurement, forcing a trade-off between dose and resolution. Our framework can explicitly model this radiation-induced evolution, enabling full-dose imaging while compensating for structural changes. Beyond microscopy, we can also handle atmospheric turbulence in long-range ptychographic imaging[52,53] by learning spatiotemporal phase distortions directly from the measurement sequence, eliminating the need for adaptive optics. For endoscopic implementations, we can model the fiber bundle modulation profile as temporal varying coded surface for synthetic aperture imaging[54,55]. Across all scales, temporal variation transforms from a fundamental limitation into useful information that enhances reconstruction quality.

Future developments could extend the reported framework in multiple directions. Integration with event-based sensors could enable microsecond temporal resolution for capturing rapid cellular dynamics. Combining our spatial-temporal factorization with multi-wavelength measurements could provide spectroscopic information alongside morphological data[56,57]. The neural field representation naturally extends to three-dimensional tomographic reconstruction by incorporating multiple illumination angles[58,59]. Moreover, the compressed representation facilitates machine learning analysis directly on the latent features, potentially enabling real-time classification of cellular behaviours or pharmaceutical quality metrics. As sensors approach fundamental limits while computational power continues exponential growth, such approaches will increasingly define the frontier where algorithm sophistication overcomes physical constraints.

## Methods

**Video-rate gigapixel ptychographic sensing system.** The video-rate ptychographic system is a compact, handheld device with the coded surface fabricated directly on top of the image sensor (onsemi AR2020, 5120 by 3840 pixels, 1.4 μm pixel size, 30 fps). The optical scheme is shown in Supplementary Fig. S11, where a 405-nm laser diode (Thorlabs, LP405-SF10) illuminates the sample from with approximately 10-mW power, providing coherent illumination across the entire centimeter-scale field of view. The coded surface consists of a disorder-engineering pattern with sub-micron intensity and phase scatterers. The coded surface pattern is pre-calibrated using a static blood smear as a reference object through joint recovery with the ePIE algorithm. The sample is positioned at a distance of 0.2-2 mm above this surface. For dynamic imaging, the coded sensor translates continuously beneath the sample using magnetic actuation from voice coil actuators, achieving acquisition rates up to the sensor framerate of 30 fps. The lateral positions $(x_t, y_t)$ at each time point are tracked using a static fiducial marker placed at the corner of the field of view. Cross-correlation between the fiducial's diffraction pattern in consecutive frames provides sub-pixel registration accuracy for distinguishing sensor motion from sample dynamics. In Supplementary Note 1, we discussed the calibration process of the coded surface and the positional tracking of sensor motion. In Supplementary Note 2, we discussed the imaging model of our ptychographic system and conventional phase retrieval algorithms.

**Space-time neural field representations.** The space-time neural field framework decomposes the complex optical field into low-rank spatial and temporal components, addressing the fundamental challenge of jointly reconstructing all time points while maintaining computational tractability. As demonstrated in Supplementary Fig. S1, we employ separate neural field generators for real and imaginary components rather than amplitude-phase representation, avoiding phase-wrap discontinuities that would otherwise cause convergence failures at crystal boundaries and regions with rapid phase variations.

The spatial feature representation leverages multi-resolution hash encoding (Supplementary Note 3), a technique that captures image details at different scales simultaneously. The encoding employs $L$ hierarchical resolution levels, starting from a coarse pixel grid and progressively refining by a factor of 1.3 at each level. For example, the first level covers 128×128 pixels, the second level 166×166 pixels, the third 216×216, and continues to the $L^{th}$ level. This multi-scale hierarchy ensures that both broad structural patterns and fine details are captured effectively. At each resolution level, the hash encoding extracts $F$ feature values per spatial location. Across all $L$ levels, this yields $F \cdot L$-dimensional feature vectors that serve as compact spatial descriptors. These $F \cdot L$ numbers encode everything from coarse object boundaries to finest interference patterns, providing the neural network with rich multi-scale information to reconstruct the complex field. Supplementary Fig. S12 shows the $F \cdot L$ feature channels with real-imaginary splitting, enables efficient representation of dynamic processes while maintaining spatial resolution. Supplementary Fig. S13 compares reconstruction performance with different activation functions in the neural field framework.

The hash table implementation as detailed in Supplementary Note 3 provides high storage efficiency for compressing the space-time volume. Rather than storing features at every pixel location which would require billions of parameters for gigapixel images, the hash encoding uses a lookup table with up to $2^{26}$ entries that are shared across all spatial positions through a hashing function. This achieves up to 1000-fold compression while maintaining high spatial resolution in reconstruction. The system maps normalized spatial coordinates $(x, y) \in [0,1]^2$ to these stored features through hash table lookups at surrounding grid vertices, with bilinear interpolation ensuring smooth transitions between discrete grid points (Supplementary Note 3).

For temporal representation, we employ $T$ learnable feature vectors $\tau_i \in \mathbb{R}^{32}$ ($i = 1,2,...T$) that capture time-varying patterns across the acquisition sequence. Linear interpolation between adjacent temporal features enables continuous reconstruction at arbitrary time points. These temporal features combine with the spatial features through element-wise multiplication (Hadamard product). The resulting fused features are decoded by dual MLPs. The dual-MLP architecture separately generates real and imaginary components of the complex field, avoiding phase discontinuities inherent in amplitude-phase representations. Each MLP output is reshaped from flattened feature vectors to 2D spatial grids matching the target reconstruction dimensions, producing the final complex-valued optical field at each time point.

**Reconstruction via gradient-domain loss.** The neural field parameters are optimized using a physics-informed loss operating in the gradient domain. The forward model follows the physical imaging process: the complex object field first propagates to the coded surface plane through angular spectrum propagation over distance $d_1$, modulates with the shifted coded surface pattern based on tracked sensor positions, then propagates to the detector plane over distance $d_2$. The final intensity is down-sampled by the magnification factor to match the raw measurement dimensions. Rather than directly comparing measured and predicted intensities, the adopted gradient-domain loss operates on spatial derivatives of amplitude fields, computing gradients of the square root of intensities. This formulation provides superior convergence demonstrated in Supplementary Fig. S5-S6.

The reconstruction process employs the Adam optimizer with an initial learning rate of $10^{-3}$, typically converging within 10 epochs. The key to robust performance lies in exploiting temporal correlations across the entire image sequence -- transforming what would be an underdetermined single-frame problem into a well-constrained global optimization. While conventional iterative methods (detailed in Supplementary Note 2) produce artifacts and often fail to converge entirely, our spatiotemporal approach maintains stable reconstruction even for challenging dynamic samples in Supplementary Fig. S5-S6 and Movie 4. Supplementary Fig. S14 further demonstrates this advantage using bone marrow mesenchymal stem cells, where our framework successfully tracks multiple cells through dynamic morphological changes while maintaining phase continuity and structural fidelity.

The framework also demonstrates robustness to temporal sampling rates ranging from 30 to 3 fps (Supplementary Fig. S15). Different frame rates are achieved using a multi-frame strategy where consecutive raw diffraction images share temporal coordinates: for 30 fps, each raw image has a unique temporal coordinate; for lower rates, multiple images share coordinates (every 2 images for 15 fps, every 5 for 6 fps, every 10 for 3 fps). Our framework enables robust recovery across all temporal sampling conditions.

**Mesoscale wound healing monitoring within incubator.** Bone marrow mesenchymal stem cells (RASMX-01001, Cyagen) were cultured in a specialized growth medium (RASMX-90011, Cyagen) supplemented with 10% FBS. For wound healing assays, cells were seeded in standard culture dishes at a density of $2 \times 10^5$ cells/cm$^2$ and grown to 90% confluence. Linear wounds were created using a sterile pipette tip by scraping through the cell monolayer. The culture dishes were then gently washed with PBS to remove floating cells and replenished with fresh medium. The entire ptychography system was positioned inside a standard incubator (37°C, 5% $CO_2$), enabling continuous monitoring without sample transfer -- eliminating mechanical perturbations and contamination risks that compromise traditional time-lapse microscopy of wound healing dynamics.

**Dry-mass tracking of bacterial growth.** We used the *E. coli* ATCC 25922 strain for the bacterial growth experiment. The bacteria were cultured in fresh Mueller-Hinton broth. In Fig. 4b, the dry mass density at each pixel is calculated as $\rho(x,y) = (\lambda/2\pi\gamma)\phi(x,y)$, where $\lambda$ is the center wavelength, $\gamma$ is the average refractive increment of protein (0.2 mL/g)[60], and $\phi(x,y)$ is the phase recovered using the video-rate ptychography system.

**Transdermal microneedle patch fabrication and characterization.** Microneedle patches were fabricated through a multi-step molding process. Silicon master molds with pyramidal cavities were first created via two-photon polymerization (Nanoscribe), achieving sub-micron feature resolution. These masters were replicated in polydimethylsiloxane (PDMS) to produce flexible, reusable negative molds. For the biodegradable microneedles, we selected polylactic-co-glycolic acid (PLGA, 50:50, acid-terminated, 15 kDa), which degrades via hydrolysis in physiological conditions. PLGA solution (20% w/v in acetone) was cast, dried, and compression-molded into the PDMS cavities at 60°C under vacuum. The final patches contained approximately 100 microneedles per 1×1 $cm^2$ area, with each needle designed to be 700 μm in height with a base diameter of 300 μm. These dimensions were specifically chosen to ensure effective penetration of the stratum corneum while maintaining sufficient mechanical strength for insertion. For 3D characterization of the fabricated device, the microneedle patches were imaged to obtain quantitative phase maps across the entire device area. The recovered phase images were then unwrapped, and the physical height was calculated using the equation $h(x,y) = (\lambda/2\pi\Delta n)\phi(x,y)$, where $\phi(x,y)$ is the unwrapped phase, and $\Delta n$ is the refractive index difference of the microneedle and the surrounding environment. The reconstructed 3D profiles provided comprehensive geometric information including needle height, tip sharpness, and array uniformity.

**Validation with extreme ultraviolet ptychography.** To demonstrate wavelength-agnostic performance, we validated our framework on extreme ultraviolet ptychography where probe instabilities challenge conventional reconstruction. The same low-rank temporal decomposition that captures sample dynamics in visible light was adapted to model time-varying probe beam $P_t(x,y)$ while keeping the object static (implementation details in Supplementary Note 4). We benchmarked our approach against conventional OPR methods with ePIE, mPIE, quasi-Newton, and least-squares algorithms, under progressively challenging overlap conditions of 24.4%, 10.3%, and 1.2% (Supplementary Figs. S18-S22). While OPR methods decompose probe variations into fixed orthogonal spatial modes, our framework models probe dynamics as continuous temporal functions. At the overlap of 24.4%, conventional OPR methods exhibit artifacts and distortions in the recovered Siemens star test pattern. As overlap decreases to 10.3% and especially at the extreme 1.2% overlap, OPR methods produce heavily degraded reconstructions with severe artifacts. In contrast, our space-time approach maintains recognizable feature recovery across all overlap conditions by leveraging temporal coherence in probe evolution. These results demonstrate that modeling temporal dynamics through neural fields provides a fundamentally more robust approach than static modal decomposition, extending the operational regime of ptychography to previously inaccessible low-overlap conditions critical for low dosage imaging and sensing applications.

**Data availability**
The experimental datasets that accompany the code implementations are available in Supplementary Note 5 with Zenodo repository links.

**Code availability**
The code implementations for space-time neural field reconstruction and associated analysis scripts are available in Supplementary Note 5 with Zenodo repository links.

## Acknowledgments

This work was partially supported by the Department of Energy SC0025582 and the National Institute of Health R01-EB034744. The content of the article does not necessarily reflect the position or policy of the US government, and no official endorsement should be inferred. Q. Z. acknowledges the support of G. E. fellowship.

## Author contributions

G. Z. conceived the original concept and supervised the project. R. W. performed the experiments. Q. Z. and Z. H. developed the space-time neural field framework. R. W., Q. Z., Z. H., and Q. M. prepared the display items and Supplementary Movies. Q. Z. prepared the Supplementary Information. F. H. and T. D. N. prepared the stem cell and microneedle experiments. T. W., S. J., L. Y., L. J., L. S., D. G., M. L., C. A., A. B., and D. B. participated in the discussion and interpretation of the results. Q. Z. and G. Z. wrote the manuscript with the input from all authors.

## Competing interests

G. Z. is a named inventor of a related patent application. T. D. N. has conflict of interest with PiezoBioMembrane Inc. and SingleTimeMicroneedles Inc. Other authors declare no competing interests.

## Supplementary information

**Supplementary Figures S1-S22:**

Supplementary Fig. S1 | Comparison between real-imaginary and amplitude-phase representations in neural field reconstruction.
Supplementary Fig. S2 | In-situ video-rate imaging of snowflake melting dynamics.
Supplementary Fig. S3 | Gigapixel-scale amplitude reconstruction of $Na_2CO_3$ crystal revealing sub-micron features over centimeter-scale field of view.
Supplementary Fig. S4 | Gigapixel-scale quantitative phase imaging of $Na_2CO_3$ crystal.
Supplementary Fig. S5 | Algorithmic performance comparison via the $Na_2CO_3$ crystallization process.
Supplementary Fig. S6 | Algorithmic performance comparison via the microneedle dissolution process.
Supplementary Fig. S7 | Post-measurement digital refocusing of thick samples.
Supplementary Fig. S8 | Post-measurement digital refocusing for imaging cell necrosis dynamics.
Supplementary Fig. S9 | Mesoscale in-situ imaging of bacterial microcolonies.
Supplementary Fig. S10 | Mesoscale characterization of transdermal microneedle patch.
Supplementary Fig. S11 | Schematic of the optical layout and principle of video-rate ptychography.
Supplementary Fig. S12 | Multi-resolution hash-encoded spatial feature maps.
Supplementary Fig. S13 | Comparison of activation functions in video-rate ptychography.
Supplementary Fig. S14 | Comparison of dynamic reconstruction methods for bone marrow mesenchymal stem cells.
Supplementary Fig. S15 | Robustness of video-rate ptychography to temporal sampling.
Supplementary Fig. S16 | Multi-resolution hash encoding for spatial feature representation.
Supplementary Fig. S17 | Optimization of hash encoding parameters for spatial feature representation.
Supplementary Fig. S18 | Effective EUV probe beam size determination from video-rate ptychography recovery.
Supplementary Fig. S19 | Video-rate ptychography reconstruction under different overlaps.
Supplementary Fig. S20 | Comparison of orthogonal probe relaxation methods with video-rate ptychography under 24.4% overlap.
Supplementary Fig. S21 | Comparison of orthogonal probe relaxation methods with video-rate ptychography under 10.3% overlap.
Supplementary Fig. S22 | Comparison of orthogonal probe relaxation methods with video-rate ptychography under 1.2% overlap.

**Supplementary Notes 1-5:**

Supplementary Note 1 | Experimental calibration of video-rate ptychographic system

Supplementary Note 2 | Temporal reconstruction strategy for conventional phase retrieval algorithms
Supplementary Note 3 | Hash encoding and data compression
Supplementary Note 4 | EUV ptychography with time-varying probe
Supplementary Note 5 | Open-source dataset description for video-rate ptychography

**Supplementary Movies 1-12:**
Supplementary Movie 1 | Temporal phase imaging of large snowflake melting dynamics.
Supplementary Movie 2 | Temporal phase imaging of small snowflakes melting dynamics.
Supplementary Movie 3 | Gigapixel-scale phase imaging of $Na_2CO_3$ crystallization dynamics.
Supplementary Movie 4 | Comparison of time-resolved recovery using different reconstruction methods.
Supplementary Movie 5 | Time-lapse monitoring of bone marrow mesenchymal stem cell division with digital refocusing.
Supplementary Movie 6 | Time-lapse monitoring of intercellular communication among bone marrow mesenchymal stem cells.
Supplementary Movie 7 | Time-lapse monitoring of bone marrow mesenchymal stem cell necrosis with dynamic focal plane correction.
Supplementary Movie 8 | Post-measurement digital refocusing of transdermal microneedle patch.
Supplementary Movie 9 | Dynamic monitoring of *E. coli* proliferation and antibiotic-induced filamentation in liquid culture.
Supplementary Movie 10 | Real-time phase imaging of *E. coli* microcolony growth on agar plate.
Supplementary Movie 11 | Continuous monitoring of microneedle dissolution dynamics.
Supplementary Movie 12 | Temporal evolution of EUV probe beam recovered by video-rate ptychography.